\documentclass[twocolumn]{aastex631}

\shorttitle{Mass transfer in symbiotic systems}
\shortauthors{I. B. Vathachira et al}

\graphicspath{{./}{figures/}}

\begin{document}

\title{Exploring mass transfer mechanisms in symbiotic systems}

\author[0009-0003-3561-5961]{Irin Babu Vathachira}
\affiliation{Department of Physics, Ariel University, Ariel, POB 3, 4070000, Israel}

\author[0000-0002-0023-0485]{Yael Hillman}
\affiliation{Department of Physics, Azrieli College of Engineering Jerusalem, Israel}
\affiliation{Department of Physics, Technion -- Israel Institute of Technology, Haifa 3200003, Israel}

\author[0000-0002-7840-0181]{Amit Kashi}
\affiliation{Department of Physics, Ariel University, Ariel, POB 3, 4070000, Israel}
\affiliation{Astrophysics, Geophysics and Space Science (AGASS) Center, Ariel University, Ariel, 4070000, Israel}

\footnotetext{\href{mailto:kashi@ariel.ac.il}{Amit Kashi: kashi@ariel.ac.il}}

\begin{abstract}

We define two regimes of the parameter space of symbiotic systems based on the dominant mass transfer mechanism. A wide range of white dwarf (WD) mass, donor mass, and donor radius combinations are explored to determine the separation, for each parameter combination, below which wind Roche-lobe overflow (WRLOF) will be the dominant mass transfer mechanism. The underlying concept is the premise that the wind accelerates. If it reaches the Roche-lobe before attaining sufficient velocity to escape, it will be trapped, and gravitationally focused through the inner Lagrangian point towards the accreting WD. However, if the wind succeeds in attaining the required velocity to escape from the donor's Roche-lobe, it will disperse isotropically, and the dominant mass transfer mechanism will be 
the Bondi-Hoyle-Lyttleton (BHL) prescription in which only a fraction of the wind will be accreted onto the WD. We present, these two regimes of the four dimensional parameter space, covering 375 different parameter combinations.

\end{abstract}

\keywords{accretion, accretion disks--- binaries: close --- binaries: symbiotic--- cataclysmic variables, novae---stars: AGB and post-AGB}

\section{Introduction} \label{sec:intro}

A nova occurs when a critical amount of hydrogen rich matter is accumulated on the surface of a white dwarf (WD), leading to a bright eruption. This process is initiated by the transfer of mass from its binary companion to the WD's surface. This results in the gradual increase of the sub-surface pressure and temperature \cite[]{1978ARA&A..16..171G,2011arXiv1111.4941B}, eventually triggering the ignition of the accreted hydrogen \cite[]{1987Ap&SS.131..379S,2016PASP..128e1001S}. The degenerate conditions inhibit cooling, thus resulting in a rapid acceleration of the nuclear burning rate, i.e., a thermonuclear runaway (TNR) \cite[]{1971MNRAS.152..307S,1972ApJ...176..169S,1978A&A....62..339P,1978ApJ...226..186S,1986ApJ...310..222P,2016PASP..128e1001S,2022ApJ...938...31S}.

Novae occur in binary systems that comprise a WD, and either a less evolved non-degenerate companion, i.e., cataclysmic variables (CVs), \cite[]{2003cvs..book.....W} or evolved giants, i.e., symbiotic systems \cite[]{1983ApJ...273..280K,1998IBVS.4571....1C,2010arXiv1011.5657M,2007NewAR..51..524P,2012PASP..124.1262B,2018A&A...612A.118L,2021MNRAS.501..201H,2023MNRAS.522.6102T}. The characteristics of a nova are contingent primarily upon the mass of the WD and the accretion rate \cite[]{1984ApJ...281..367P,1994ApJ...424..319K,1995ApJ...445..789P,2005ApJ...623..398Y,2012BaltA..21...76S}, while the companion's mass and stellar type as well as the binary separation of the system, determine the system's mass transfer rate \cite[]{1988A&A...202...93R,2007BaltA..16...26P,2018MNRAS.473..747C,2021MNRAS.501..201H,2024MNRAS.532.1421T}. In systems with Mira-type variable star donors, abrupt changes in the mass loss rate during the donor's evolution can significantly impact the recurrence time between eruptions. This is evident in RX Pup, where a threefold increase in the mass transfer rate shortened the recurrence time by 4.5 times \cite[]{2024ApJ...972L..14I}.

Novae are most commonly known to occur in CVs, where the companion is a red dwarf (RD) that fills its Roche lobe (RL), initiating the transfer of mass to the WD through the inner Lagrangian point ($\rm L_1$). This process of mass transfer through $\rm L_1$ is called Roche lobe overflow (RLOF) \citep{1941ApJ....93..133K,1971ARA&A...9..183P,1987Ap&SS.131..379S}, and the accretion rate in this case is exponentially dependant on the extent of the overflow ($ R_{\rm RL}-R_{\rm RD}$, where $ R_{\rm RL}$ and $ R_{\rm RD}$ are the radii of the RL and the RD respectively)
\citep{1988A&A...202...93R,2020NatAs...4..886H,2021MNRAS.505.3260H}, while the RL radius is directly linked to the separation of the system \cite[][see Equation \ref{a_orig} below]{1983ApJ...268..368E}. 

In symbiotic systems, a red giant (RGB) or asymptotic giant branch (AGB) companion loses mass via wind, and while for the former, this rate typically consists of steady flows, the latter may experience substantial fluctuations in the wind rate, resulting from thermal pulses \cite[]{1983A&A...127...73B,1993ApJ...413..641V,dorfi1998agb}. These thermal pulses are the alternating nuclear burning of hydrogen and helium layers in the AGB's outer envelope culminating as a series of contractions and expansions, causing fluctuations in radiation pressure that lead to rapid changes in both radius and wind velocity \cite[]{1975ApJ...196..525I,1978PASJ...30..467S,doi:10.1146/annurev.aa.21.090183.001415}.
This leads to a significant rate of mass loss from the AGB star, usually in the range  $10^{-7}-10^{-4}\mathrm{\mathrm{M_\odot yr^{-1}}}$ \cite[]{2005ARA&A..43..435H,2009ASPC..414....3H,2003IAUS..210P.A13A}. After the conclusion of this mass loss phase, the remaining material contracts towards the beginning of the WD phase \cite[]{2009ASPC..414....3H}.

In principle, RLOF is a viable mechanism of mass transfer in symbiotic systems as well, and there is indication of this in certain S-type  systems \cite[]{2007BaltA..16....1M,2008ASPC..401...42M}. Moreover, there is observational evidence supporting the idea that RLOF is the most probable mass transfer mechanism in S-type systems with orbital periods of order a few years \cite[]{2010arXiv1011.5657M}. 

In D-type systems mass transfer occurs mainly by wind \cite[]{2007BaltA..16....1M,2008ASPC..401...42M} and a portion of this wind may be captured by the WD's gravity. If the system is sufficiently wide, the matter may be assumed to be accreted via the Bondi-Hoyle-Lyttleton (BHL) mechanism \cite[]{1944MNRAS.104..273B}.
The BHL method describes the process through which a star moving within a cloud of gas may gravitationally capture mass from the cloud, to be accumulated onto the stars surface. This mechanism may be applied to a WD in a symbiotic system, where the donor is losing mass via an (assumed to be) isotropic wind, and the WD, being enveloped by this wind, captures a fraction of it, while the rest is expelled from the system \cite[]{1939PCPS...35..405H,1944MNRAS.104..273B,1952MNRAS.112..195B,2004NewAR..48..843E}. Works by \cite{2021MNRAS.501..201H} and \cite{2024MNRAS.527.4806V} demonstrate the dependence of the binary separation on the rate that the WD accretes matter from the donor's wind (i.e., the accretion rate, $ \dot{M}_{\rm acc}$), assuming the dominant mass transfer mechanism to be the BHL mechanism described above. These authors obtained higher accretion rates for models with smaller binary separations which favoured the growth of the WD mass.

An important factor that influences the mass transfer rate in symbiotic systems, is the wind velocity. When considering the BHL mechanism, the giant's RL radius is regarded as irrelevant since the wind is assumed to have a velocity high enough to escape the giant's RL. However, if the wind were to approach the RL's limit with a velocity that is slower than the RL escape velocity, it would be trapped in the RL and hence focused through the $\rm L_1$ point to be accreted onto the WD, i.e., \textit{gravitational focusing} of the giant's wind. This mass transfer mechanism is commonly referred to as \textit{wind Roche lobe overflow} (WRLOF) \cite[]{2007BaltA..16...26P}. When the wind is expelled from the giant's surface, its velocity is relatively low, of order $\sim5\rm km/s$ \cite[]{2003A&A...409..715W,2004JBAA..114..168P,2019A&A...626A..68S,2021A&A...653A..25M}, which is typically less than the velocity required to escape the giant's RL \cite[]{2013A&A...552A..26A}. 
However, the wind accelerates as shock waves, arising from stellar pulsation and vigorous convective motions, traverses through the expelled wind. These waves push the gas much further, prompting the creation of dust particles. These particles then collide, gather momentum and experience additional acceleration \cite[]{2009ASPC..414....3H,2016csss.confE...5W,2018A&ARv..26....1H}.
 If this mass reaches the AGB's RL limit before accelerating up to the escape velocity, then it will be trapped, and consequently focused through the inner lagrangian point, L1, towards the WD. 
  This means that there is a limit to the distance between the donor radius and its RL radius, below which the wind will be trapped and transferred to the WD via the WRLOF mechanism. On the other hand, if this distance is larger than the limit described above, the AGB's wind would manage to reach the escape velocity before approaching its RL's outer limit, thus, it will escape the system, and the dominant mechanism of mass transfer to the WD will be the BHL mechanism, as described above. The nature of these two mechanisms, dictates that the mass transfer of the former will be significantly more efficient than that of the latter, since the latter results in most of the wind being lost from the system.

The concept of WRLOF, or gravitational focusing of wind, dates back to the study by \cite{1997ApJ...487..809H}, where they highlighted the intriguing possibility of two distinct mass transfer mechanisms coexisting within the same system depending on their orbital separation. In their study, they considered eccentric models of the Egg Nebula, where the mass lost by the spherically symmetric wind of the giant donor was regarded as the primary contributor to the nebula's matter. 
The type of mass transfer and accretion depended on the wind's ability to attain escape velocity \cite[]{1981ApJ...249..572M,1988ApJ...329..299B,1993ApJ...413..641V}. These authors have shown that in cases of eccentric orbits, the donor may experience two types of mass loss mechanisms: (i) when the separation is small, i.e., in the vicinity of periastron, the expelled wind does not exceed the velocity required to escape from its RL, thus, it is focused towards the equatorial plane of the accretor; (ii) when the separation is large, i.e., in the vicinity of apastron, the mass loss occurs isotropically because the giant's wind would have attained the velocity required to escape from its RL, thus, the mass transfer to the accretor occurs via the BHL mechanism. The reason that the separation governs the dominant mass transfer mechanism is because the separation is proportional to the giant's RL radius \cite[]{1983ApJ...268..368E}. Therefore, a smaller separation indicates a smaller RL radius, thus, the wind reaches the outer RL limit after traveling a shorter distance, preventing the wind from attaining escape velocity. In contrast, a larger separation results in a larger RL radius, yielding a longer distance between the giant's surface and its RL limit, thus, allowing the wind to accelerate beyond the velocity required to escape the system isotropically.

\cite{2007BaltA..16...26P} compared the mass transfer efficiency of Mira-type symbiotic stars by taking into account both WRLOF and BHL mechanisms using a hydrodynamical code. 
The study was conducted using the same stellar masses with only their separation varying to examine the behavior of the mass transfer. They found that for smaller separations, the dust formation and escape velocity were attained beyond the giant's RL radius, causing the donor's wind to be focused towards the equatorial plane of the WD through the $\rm L_{1}$ point, i.e., the dominant mass transfer mechanism in this case is WRLOF. In contrast, larger separations resulted in a much larger RL radius, allowing dust formation and escape velocity to occur within the RL. Consequently, the wind escaped isotropically from the donor, indicating the BHL mass transfer mechanism. Their results showed, for their given stellar parameters and wind rates, that in the case of WRLOF, almost 50\% of the matter was transferred to the accretor as apposed to only a few percent of the wind that is accreted onto the WD for their BHL cases. Consequently, the mass transfer rate following the WRLOF mechanism was at least ten times higher than the corresponding values observed in the BHL scenario.

An additional study conducted by \cite{2018MNRAS.473..747C} involved simulations aimed at exploring the mass transfer mechanism of symbiotic systems hosting AGB donors. Their 3D hydrodynamical simulations investigated how orbital separation, stellar mass ratios, and accretion processes influence the evolution of the system. These authors conducted simulations using five different models, with binary separations in the range $640-2150{R_\odot}$ ($3-10$ AU). Their findings demonstrated that for smaller separations ($<6$ AU), the systems experienced mass transfer via the WRLOF mechanism while some of the matter is not accreted but rather lost through the outer Lagrangian point, $\rm L_2$, causing the binary separation to decrease. In contrast, systems with greater separations showed signs of wind accretion via the BHL prescription, leading to a widening of the binary.

\begin{figure}
 %\fbox
{    \includegraphics[trim={5.0cm 0.5cm 8.5cm 1.5cm},clip,width=0.99\columnwidth]{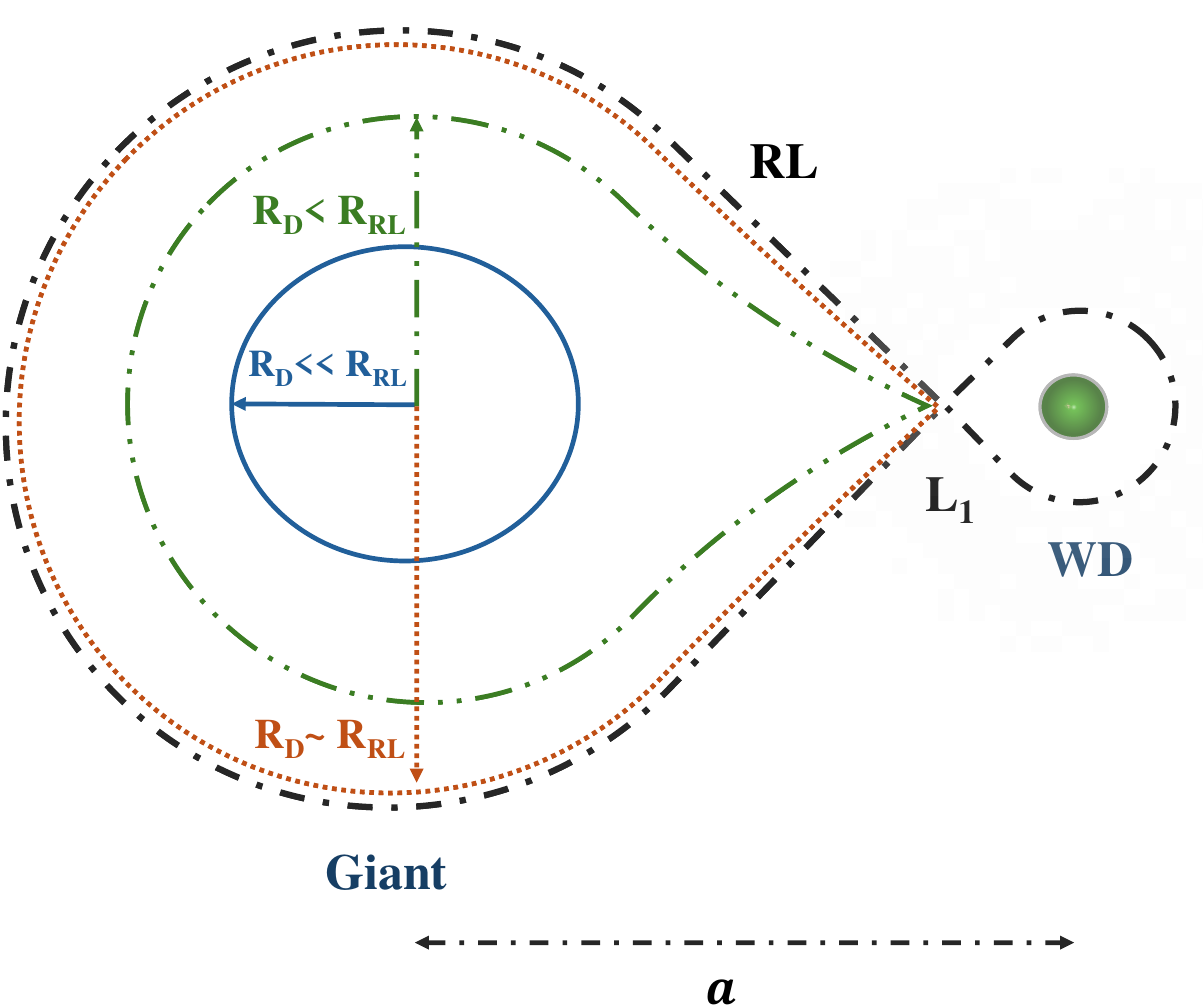}}
   \caption{A schematic representation of a binary system consisting of a given $ M_{\rm WD}$, $M_{\rm D}$, and $a$. The corresponding Roche lobe (RL) radius, calculated using the Eggleton prescription, is marked as $ RL$ and outlined with a dash-dotted black line. The three possible regimes for the donor radius, $ R_{\rm D}$, are marked as follows: (1) $ R_D \sim R_{\rm RL}$ in dotted brown (where the donor radius is approximately the same as the RL radius, and mass transfer occurs via the RLOF mechanism); (2) $ R_{\rm D} < R_{\rm RL}$ in dash-dotted green (where the wind does not reach escape velocity before reaching the RL radius and is transferred to the WD through $\rm L_1$, with mass transfer occurring via the WRLOF mechanism); and (3) $ R_{\rm D} << R_{\rm RL}$ in solid blue (where the wind attains escape velocity before reaching the RL radius and escapes, with mass transfer occurring via the BHL mechanism).}
   \label{WRLOFpic}
\end{figure}

We show a schematic presentation of such a binary system in Figure \ref{WRLOFpic}, demonstrating how, for a given donor mass, WD mass and binary separation, the radius of the giant donor and its RL radius may determine the mass transfer mechanism. We illustrate three basic formations that yield three basic mass transfer mechanisms. The first is for a case where the radius of the giant, $ R_{\rm D}$, is approximately equal to its RL radius, $ R_{\rm RL}$, (i.e., $ R_{\rm D}\sim R_{\rm RL}$), resulting in the giant filling its RL and transferring mass by RLOF. The second case is that of the giant's radius being smaller than its RL radius ($ R_{\rm D}< R_{\rm RL}$) --- but not small enough to provide the wind with a travelling distance sufficient for accelerating beyond the RL escape velocity --- the dominant mass transfer mechanism will be WRLOF. In this case, the wind from the donor does not reach escape velocity before reaching the RL radius, $ v_{\rm wind}< v_{\rm esc,RL}$, and is transferred to the WD through the $\rm L_1$ point. The third case is of the giant's radius being \textit{much} smaller than the RL radius ($ R_{\rm D}<< R_{\rm RL}$) --- thus the wind will attain the velocity required for escaping the giant's RL before reaching the RL's outer limit, $ v_{\rm wind}> v_{\rm esc,RL}$, and it will escape from the system, with the dominant mass transfer mechanism being the BHL accretion scenario. In this case the mass transfer to the WD will be inefficient because only a small fraction of the wind is captured by the WD. 

Under certain circumstances, the dominant mass transfer mechanism in a binary system with a given initial separation, '$a$', can switch from one type to another. It is highly plausible that this can occur in systems with AGB donors, because their radii can rapidly change over the system's AGB phase\footnote{Such shifting from one mechanism to another may also occur in eccentric systems following the same principal, as has been demonstrated for the Egg nebula \cite[]{1997ApJ...487..809H}.}.
The determination of the mass transfer mechanism in a symbiotic system depends on the masses of the WD and the giant donor, the radius of the donor, the binary separation and the wind's velocity. 
In the present work we explore the entire feasible ranges of the four free stellar parameters ($ M_{\rm WD}, M_{\rm D}, R_{\rm D}$ and $a$), and define limits that differentiate between two parameter regimes. One that is governed by WRLOF and the other by BHL as the dominant mass transfer mechanism. 
In the following section, we describe our method of calculations and models, followed by our results in section \ref{result}. We discuss our findings and their consequences in section \ref{discussion} and present our main conclusions in section \ref{Conclusion}.

\section
{Methods: analytical approach and numerical models} \label{sec2}

\subsection{Estimates and calculations}

In accordance with the prescription outlined by \cite{1983ApJ...268..368E}, the Roche-lobe radius ($ R_{\rm RL}$) of a star in a binary system is determined by the WD mass ($ M_{\rm WD}$), the donor mass ($ M_{\rm D}$) and the binary separation ($a$), and is given as:
\begin{equation} \label{a_orig}
  R_{RL} =  \frac{0.49q^{2/3}}{0.6q^{2/3}+\ln(1+q^{1/3})}\textit{a},
\end{equation}
where if $ R_{\rm RL}$ is the RL radius of the donor, then '$ q$' is the mass ratio, ${M_{\rm D}/M_{\rm WD}}$. 

 In order to achieve WRLOF, wind is expected to blow until the limits of the RL radius, and then be transferred via the $ \rm L_1$ to the WD. For this to occur, the wind must reach the donor's $ R_{{\rm RL}}$ before attaining the velocity required to escape the RL  (\rm $v_{esc,RL}$). This way, it will be trapped in the RL and gravitationally focused through the $\mathrm{L_1}$ point to the WD. Since the wind that is expelled from the surface of a giant is accelerated \cite[]{2009ASPC..414....3H,2016csss.confE...5W,2018A&ARv..26....1H}, in order to achieve WRLOF, we must limit the distance the  wind travels, meaning that $R_{{\rm RL}}- R_{{\rm D}}$ (where $ R_{\rm D}$ is the donor radius) must be less than some value typical of the system. This may be better expressed as the ratio, $ R_{{\rm RL}}/ R_{{\rm D}}$, that needs to be limited to at least some typical value, based on the escape velocity from the RL ( $v_{\rm esc,RL}$). This is challenging, while the stellar masses and separation evolve relatively slow, the donor radius, can undergo radical changes. 
 This means that our primary goal is to determine the parameter space of systems that lead to mass transfer through the WRLOF mechanism, i.e, combinations for which the donor's wind does achieve the velocity required to escape its Roche lobe radius. Systems that reside outside this parameter space will be governed by the BHL mass transfer mechanism. 

 There are four free parameters that rule the evolution: 
 the stellar masses, the donor's radius and the binary separation. 
 For a given system comprising certain values of $\rm  M_{ WD}$, $\rm M_{ D}$ and $\rm R_{ D}$, we find the maximum possible RL radius that will trap the accelerating wind.
 This value is then used to calculate the maximum allowed separation, for which WRLOF will be the dominant mass transfer process. 
 We note, that since the donor radius may change rapidly over the lifetime of an AGB, we calculate this limit for multiple donor radii for the same AGB model.

 The limit of the $ R_{\rm RL}$ is determined via the wind velocity as follows.  
 For a donor star with radius $ R_{\rm D}$, the velocity of the wind at the surface of the giant is given as $ v_{{\rm s}}$, i.e., $ v_{\mathrm{w}}( r=R_{\mathrm{D}})=v_{\mathrm{s}}$. We take $ v_{\mathrm{s}}$ for each combination of AGB mass and radius, from the dense grid of models that was built for the self-consistent nova evolution code \cite[]{2020NatAs...4..886H,2021MNRAS.505.3260H}. This grid was produced by using the hydrostatic stellar evolution code developed by \cite{2009MNRAS.395.1857K}. The code is a self-sufficient program that uses an adaptive mesh grid approach/dynamic grid refinement technique, updating its values at each timestep by solving hydrostatic and energy equations while considering convective diffusion and mixing. It iterates through the most suitable mass-loss equations and incorporates OPAL opacities \cite[]{Iglesias1996}  accounting for a chain of nuclear reactions up to the production of magnesium. Thus, the code can simulate the evolution of stars, from the pre-main-sequence phase, subsequent stellar evolution stages, envelope loss, contraction, cooling, and ultimately transforming into a WD. Parameters, such as, mass, radius, density, effective temperature, luminosity, pressure, wind velocity, mass-loss rate, and other critical parameters are recorded throughout the simulation. In this work we use AGB donors for which the velocity of the wind as it departs from their surface, is typically of order $\sim5\rm km/s$. We denote the velocity of the wind at a far distance away from the surface of the donor as $ v_\infty$, i.e., $ v_{\mathrm{w}}( r>>R_{\mathrm{D}})=v_\infty$, where $ v_\infty$ is of order $\sim20\rm km/s$.
 
We use $ v_{\rm s}$ and $ v_\infty$ to attain the wind profile at a distance '$ r$' from the donor's surface \cite[]{1999isw..book.....L,Kashi_2009,2021MNRAS.501..201H}. Thus the wind velocity of a giant (e.g., an AGB) when it approaches its RL radius becomes (following, e.g., Eq. 4 from \cite{2021MNRAS.501..201H}):
\begin{equation}\label{windvelocity}
   v_w(r=R_{RL})=  v_s\frac{R_D}{R_{RL}}+v_\infty\left(1-\frac{R_D}{R_{RL}}\right).
\end{equation}
We use the escape velocity from the RL of the AGB star:
\begin{equation}\label{escape velocity}
     v_{\mathrm{esc}}(r=R_{RL}) = \sqrt{\frac{2GM_D}{R_{RL}}},
\end{equation}
as the maximum velocity the wind can attain, $v_{w}(r=R_{RL})\le v_{esc}(r=R_{RL})$, thus giving:
\begin{equation}\label{1}
     v_s\frac{R_D}{R_{RL}}+v_\infty\left(1-\frac{R_D}{R_{RL}}\right) \le \sqrt{\frac{2GM_D}{R_{RL}}},
\end{equation}
which may be rearranged and solved for $\rm R_{\rm RL}$. 

We initially considered both kinetic and thermal energies as products of the gravitational potential energy and found the contribution of the thermal energy to be negligible, thus it was not included in our calculations.
 
 The derived $ R_{\rm RL}(R_{\rm D},M_{\rm D},v_{\rm s},v_{\infty})$ upper limit for WRLOF is then used in Equation \ref{a_orig} to obtain the corresponding binary separation, $a_{\rm lim} (M_{\rm WD},R_{\rm D},M_{\rm D},v_{\rm s},v_{\infty})$. 
 This separation defines an upper limit to the regime within which WRLOF is possible, while for a wider separation, the BHL mechanism will dominate. We applied this calculation to a wide range of input model parameters as specified below.

\subsection{Models}

\begin{figure*}
%\hspace*{-1.0cm}
    \includegraphics[trim={0.0cm 1cm 0.0cm 0cm},clip,width=0.99\columnwidth]{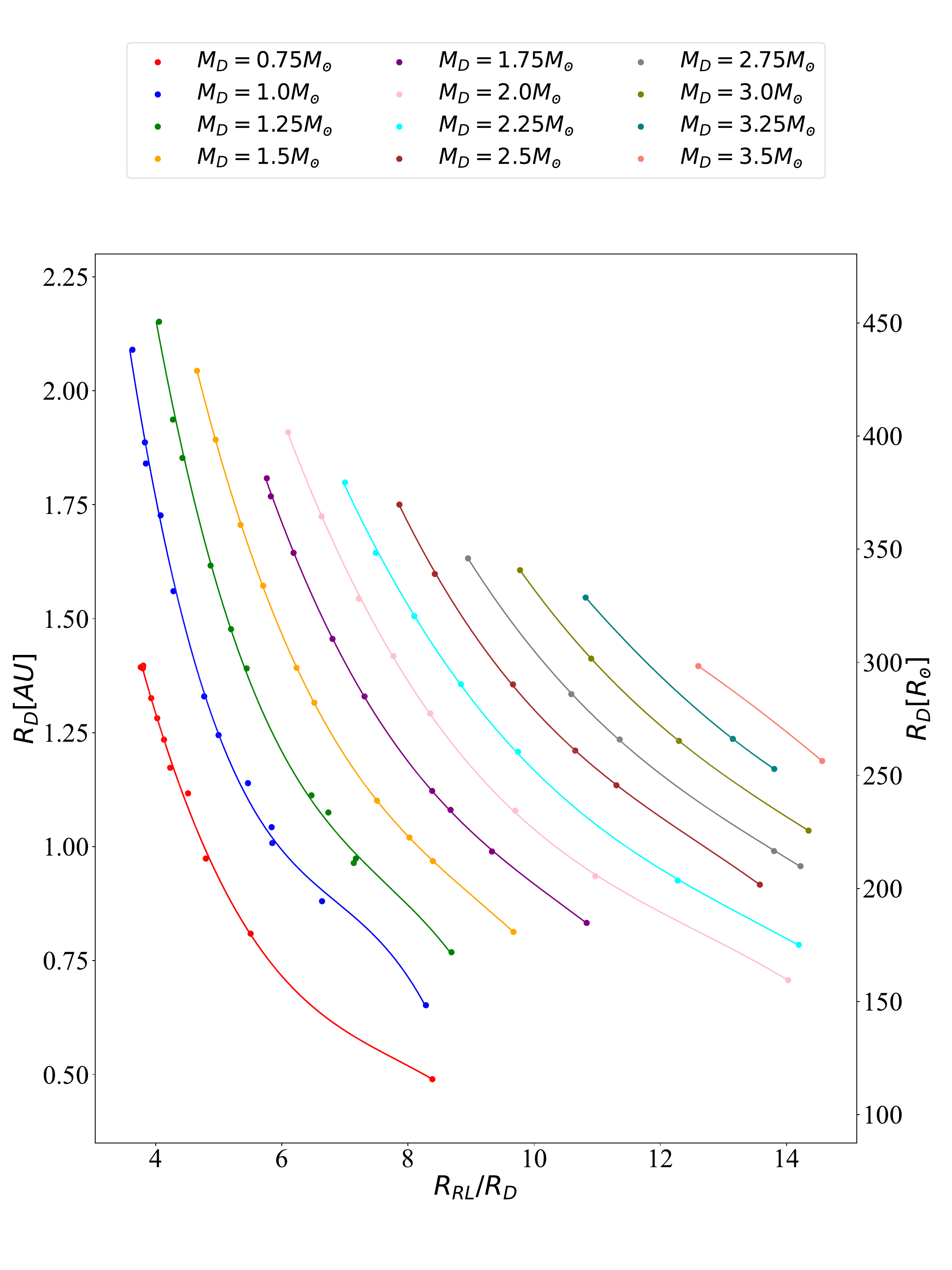}
 %   \hspace*{-0.1cm}
    \includegraphics[trim={0.0cm 1cm 0.0cm 0cm},clip,width=0.99\columnwidth]{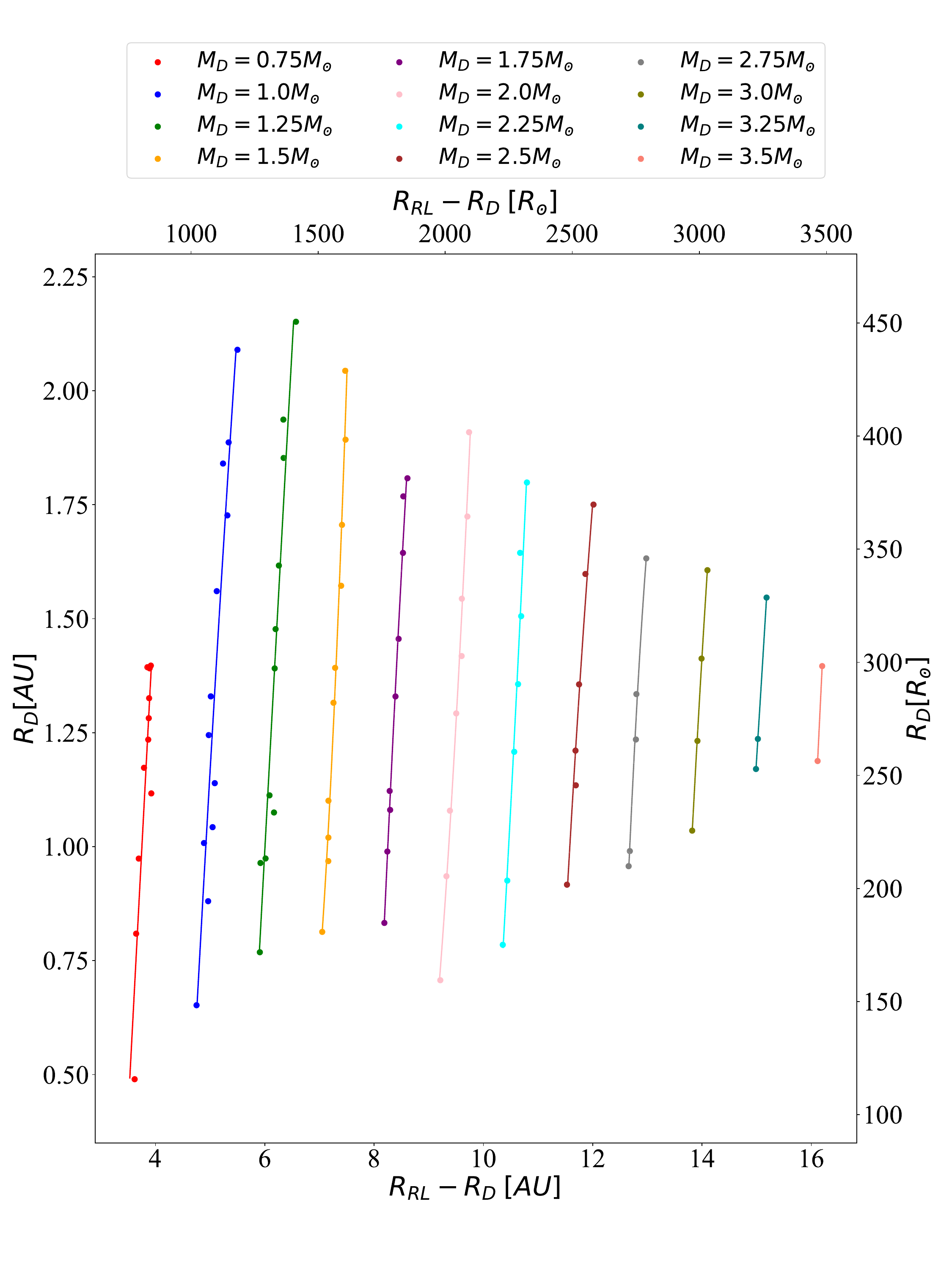}
   \caption{The donor radius $ (R_{\rm D})$ vs. the ratio of the RL radius ($ R_{\rm RL}$) to $ R_{\rm D}$ (left); and vs. the distance traveled by the wind ($ R_{\rm RL}-R_{\rm D}$) (right). Each color represents a different donor AGB mass.   Left panel shows that larger donor radii result in smaller ratios, while larger donor masses lead to larger ratios. Right panel shows that the distance the wind must travel to reach escape velocity remains nearly constant for a given donor mass but increases with donor mass due to stronger gravitational pull, requiring the wind to travel farther to escape the RL.}
   \label{Ratiorl}
\end{figure*}

For our donor AGB we used the grid of models that was built for the work in \cite{2020NatAs...4..886H}. This grid was produced by utilizing the hydrostatic stellar evolution code \cite[]{2009MNRAS.395.1857K} following the evolution of stellar models with initial masses ranging from 0.09 to 4${M_\odot}$ in intervals of 0.05${M_\odot}$. For the purpose of our study, we adopt models with initial MS masses in the range $1.0-4.0{M_\odot}$, with intervals of 0.25$ M_\odot$, resulting in a total of 13 donor AGB models. 

Throughout the evolution of each of these 13 mass losing models, we selected 12 distinct masses to include in this work, ranging from $0.75$ to $3.5{M_\odot}$. We took evolutionary points from each of the 13 models at the distinct chosen masses, and extracted the radii for these masses, yielding a total of 75 combinations of donor mass and radius ($ M_{\rm D}$ and $ R_{\rm D}$ respectively). 
We subsequently used these in Equation \ref{1} (together with the corresponding wind velocities) to obtain, for each combination of $ M_{\rm D}$ and $ R_{\rm D}$, an upper limit to the AGB’s RL radius that will allow WRLOF to be the dominant mass transfer mechanism. 
We paired each combination of $ M_{\rm D}$ and $ R_{\rm D}$, with five different WD masses ($0.6, 0.7, 1.0, 1.25$ and $ 1.4M_\odot$) to obtain 375 different combinations of $ M_{\rm D}$ and $ R_{\rm D}$ and $ M_{\rm WD}$. 
Finally, we used the two masses together with the derived upper limit to the RL radius, in Equation \ref{a_orig} to calculate an upper limit to the binary separation ($ a_{\rm lim}$). 

This process yielded 375 combinations of the four parameters, $ M_{\rm WD}$, $ M_{\rm D}$, $ R_{\rm D}$ and $ a_{\rm lim}$, that indicate the boundary between WRLOF and BHL as the dominant mass transfer mechanism. (The combinations are specified in Table \ref{tab:Table new1}.)

We note that the importance of sampling a wide range of donor radii is due to the choice of donor being a giant, specifically, an AGB, which undergoes thermal pulses. This causes the AGB's radius to undergo significant, rapid variations that may have crucial implications on the binary evolution. Specifically, it can cause the dominant mass transfer mechanism to change from BHL to WRLOF and/or vice versa.  

\section{Results}\label{result}

\begin{figure}
%\hspace*{-0.5cm}
    \includegraphics[trim={2cm 4cm 0cm 0cm},clip ,width=0.99\columnwidth]{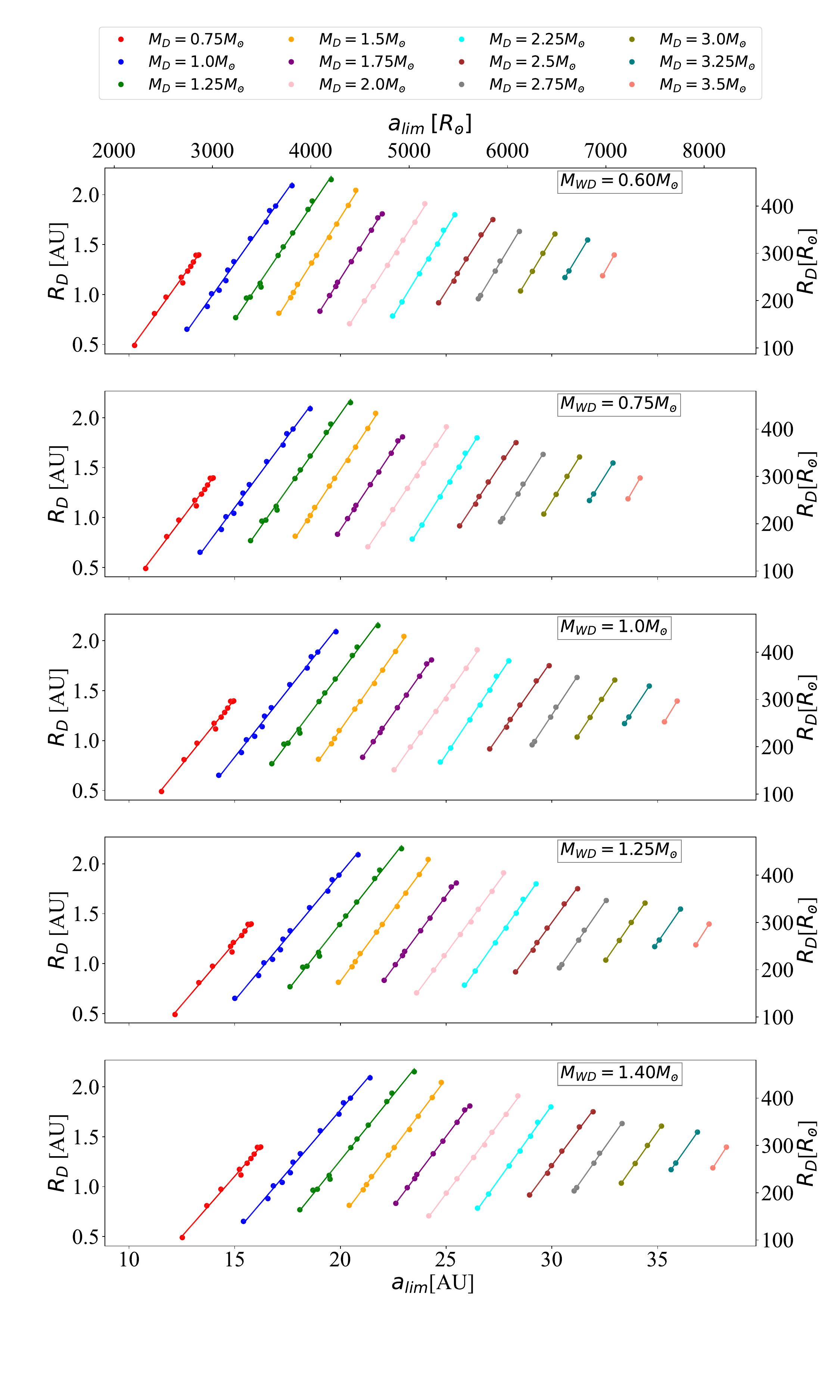}
   \caption{The limiting separation ($a_{\rm lim}$). Each colored curve represents a different AGB donor mass ($ M_{\rm D}$), and each panel corresponds to a different WD mass. A system that falls to the left of a given curve will experience WRLOF as the dominant mass transfer mechanism, while a system that falls to the right of the curve will have BHL as the dominant mass transfer mechanism. The data used for this figure is provided in Table \ref{tab:Table new1}.}
   \label{Separtion1}
\end{figure}

\begin{figure}
 %   \hspace*{-0.0cm}
    \includegraphics[trim={0.5cm 2.5cm 0.0cm 2cm},clip,width=0.99\columnwidth]{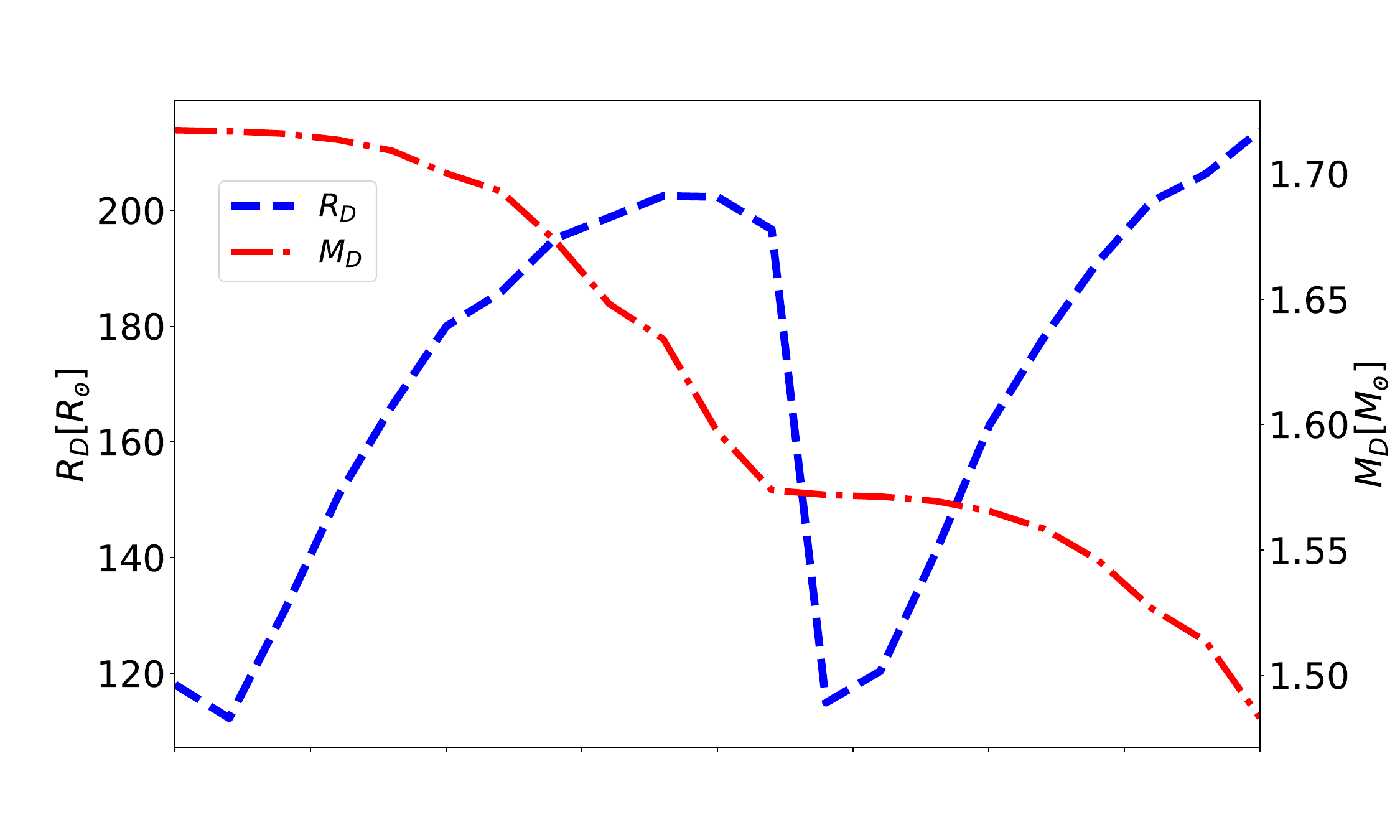}
%    \hspace*{-0.0cm}
    \includegraphics[trim={0.5cm 0cm 0.0cm 2cm},clip,width=0.99\columnwidth]{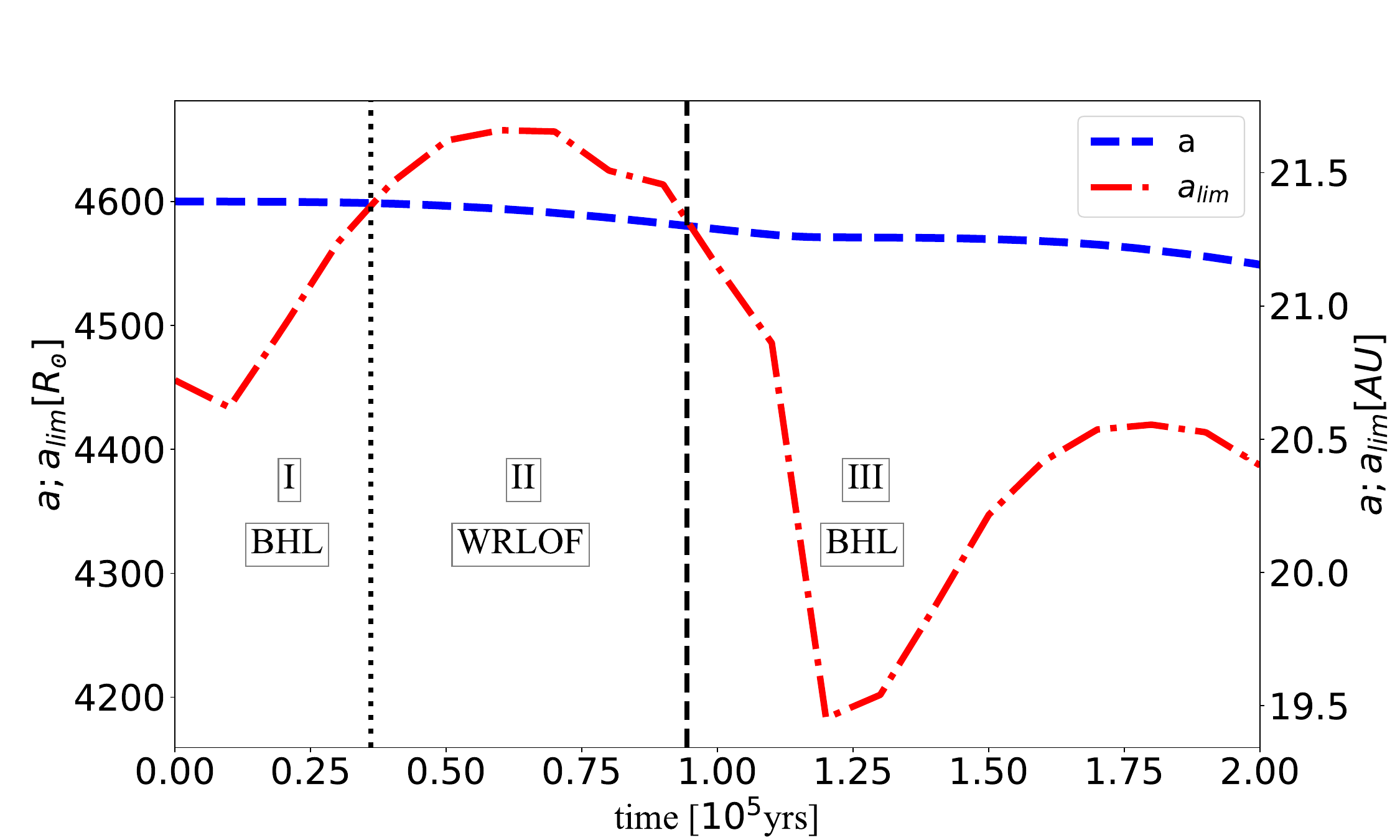}
   \caption{The upper panel in the figure shows the evolution of the mass and radius of the AGB. The bottom panel depicts the corresponding evolution of the actual binary separation, $ a$, for a 1.25 $\mathrm{M_\odot}$ WD with an AGB donor of mass 1.71 $\mathrm{M_\odot}$ and the limiting separation, $ a_{\rm lim}$, of the system. Initially, the system started with BHL accretion (region I). The vertical lines indicate instances where there was a shift in the mass transfer mechanism. The first dotted line represents the transition line to, $ a < a_{\rm lim}$, and the mechanism shifted from BHL to WRLOF (region II), continuing mass transfer by that mechanism. The dashed lined represents the instance where, $ a > a_{\rm lim}$, and the mechanism shifted from WRLOF back to BHL (region III).}
   \label{twoseven}
\end{figure}

We carried out computations as described above to attain for all our models an upper limit to the binary separation ($a_{\rm lim}$) that differentiates between parameter regimes that are dominated by the WRLOF and the BHL mass transfer mechanisms.

In Figure \ref{Ratiorl} (left) we show the donor radius ($ R_{\rm D}$) plotted against the ratio $ R_{\rm RL}/R_{\rm D}$, demonstrating that for a given donor \textit{mass}, larger \textit{radii} lead to a \textit{smaller} ratio 
and for a given donor \textit{radius}, larger donor \textit{masses} lead to a \textit{larger} ratio. This is
because the limiting $\rm R_{RL}$ is derived from the distance required for the wind to accelerate and reach the velocity required to escape the RL's outer limit. This distance will be a smaller fraction of the donor radius for larger radii and less massive stars.
This travel distance is explicitly shown in the right panel of Figure \ref{Ratiorl} illustrating that, for a given donor mass, the distance the wind must travel to reach escape velocity remains nearly constant.  This distance increases with increasing donor mass due to the fact that more massive donors exert a stronger gravitational pull, forcing the wind to travel a greater distance to accelerate sufficiently to achieve the necessary velocity to escape the RL.

From the resulting limiting RL radius, we calculated, for each parameter set, the corresponding limiting separation, $a_{\rm lim}$. This is presented in Figure \ref{Separtion1}, where each panel shows results for a different WD mass and each line represents a different donor mass. For a given set of $ M_{\rm D}$, $ R_{\rm D}$ and $ M_{\rm WD}$, separation values that are to the left of the corresponding curve, i.e., less than the limiting value, the mass transfer will occur predominantly via WRLOF, while separation values that are to the right of the corresponding curve, i.e., greater than the calculated limit, the dominant mass transfer mechanism will be the BHL mechanism.

Additionally, comparing between the curves of a given donor mass for different WD masses, it is evident that a more massive WD leads to  a higher separation limit. This may be seen by following a curve of a certain color down from the upper panel to the lower panel, demonstrating how the curve shifts to the right with increasing WD mass.  This is because for given donor RL radius and mass, the separation depends on the WD mass as dictated by Equation \ref{a_orig}, even though the calculated \textit{limiting} value of $ R_{\rm RL}$ is independent of the WD mass. 
Thus, we can deduce from Figure \ref{Separtion1} that $a_{\rm lim}$ increases with the increase of $ M_{\rm D}$, $ R_{\rm D}$ or $ M_{\rm WD}$. The illustration of these dependencies as 3-D surface plots is provided in the appendix (Figure \ref{Separtionsurf}).

Solving Equation \ref{1} yields a limiting RL radius: 
\begin{equation} \label{R_RL_lim}
  R_{\rm RL,lim} =\xi\left(\xi +\sqrt{\xi^2+2\omega R_D} \right) +\omega R_D,
\end{equation}
which can be used directly in Equation \ref{a_orig} to obtain a connection between the four key parameters ($a_{\rm lim}, M_{\rm WD},R_{\rm D}$ and $M_{\rm D}$) that can be expressed as: 
\begin{equation} \label{a_lim}
  a_{\rm lim} =R_{\rm RL,lim}  \frac{0.6q^{2/3}+\ln(1+q^{1/3})}{0.49q^{2/3}}
\end{equation}

where $\xi\approx\sqrt{500M_{\rm D}}$ is dimensionless and depends only on the donor mass that is normalized to units of $M_\odot$, $\omega$ is a dimensionless constant of order unity derived from the wind velocity, $q$ is the mass ratio as in Eggleton's prescription (see Equation \ref{a_orig}), and $R_{\rm D}$ and $a_{\rm lim}$ are given in units of $R_\odot$.
A detailed calculation of this derivation is given in Appendix \ref{Apendix}.
 
We stress that all the above are not  presentations of evolving stars over time, but rather single points in time taken as snapshots from the evolution of AGB stars.

We therefore present, as an example to the influence that a changing AGB radius can have on the mass transfer mechanism,
in Figure \ref{twoseven} a portion of the evolution of a system comprising a WD of mass $1.25 M_\odot$ and an AGB donor beginning with a mass of $\sim1.7M_\odot$ and a radius of $\sim120 R_\odot$ ($\sim0.5$AU). We arbitrarily set this point in the systems evolution as $t=0$. 
The upper panel shows the evolution of the mass and radius of the AGB with respect to time demonstrating mass loss, and rapid radius changes over the course of $\sim2\times 10^5$ years. The bottom panel shows both '$a$' and '$a_{\rm lim}$', which are the \textit{actual} binary separation of the system and the limiting value of binary separation respectively.
During the regime marked with the notation "\textbf{I}", the system's \textit{actual} separation is larger than the limit, i.e., $ a>a_{\rm lim}$, meaning that the dominant mass transfer mechanism during this epoch is the BHL scenario. The rapidly increasing radius then pushes $a_{\rm lim}$ to higher values --- higher than the actual separation, i.e., $a<a_{\rm lim}$, thus the mass transfer mechanism in this regime will be predominantly WRLOF (regime \textbf{II}). The rapidly evolving radius then decreases, shifting the limiting separation back down below the actual radius ($a>a_{\rm lim}$), thus the dominant mass transfer mechanism shifts back to the BHL scenario (regime \textbf{III}). 
Hence, for a symbiotic system, hosting an AGB donor it is plausible that the typical thermal pulses experienced by the AGB will cause the dominant mass transfer mechanism to potentially shift back and forth between BHL and WRLOF throughout its evolution.

\section{Discussion}\label{discussion}
Mass transfer in symbiotic systems can theoretically occur via three different mechanisms: RLOF, BHL accretion, or WRLOF. The dominant accretion mechanism for a given system at a given evolutionary point in time, i.e., WD mass, donor mass and donor radius, is governed by the binary separation. The transferred mass that accumulates on the surface of the WD inevitably leads to nova eruptions, while the mass transfer rate is a crucial ingredient in determining many aspects of the the outcome of the nova eruption, such as, the eruption frequency, the amount of mass retained after an eruption, the ejecta composition etc. \cite[]{1984ApJ...281..367P,1994ApJ...424..319K,1995ApJ...445..789P,2005ApJ...623..398Y,2012BaltA..21...76S,2021MNRAS.501..201H,2021MNRAS.505.3260H}.

The WRLOF models used by \cite{2012BaltA..21...88M} with $ M_{\rm WD}=0.6M_\odot$ and $ M_{\rm D}=1.0M_\odot$, yielded $ R_{\rm RL}=8.5 AU$ $(\sim 1800 R_\odot$), and $a=20\rm AU$ $(\sim 4300R_\odot)$ which is slightly above our calculated limit. The difference could possibly be the result of using different assumptions as the marker for the limit on the RL radius --- these authors assumed oxygen-rich dust and calculate the dust formation radius \cite[]{2018A&ARv..26....1H}, while we calculate the escape velocity. According to \cite{2012BaltA..21...88M} and \cite{2013A&A...552A..26A}, the difference between WRLOF and BHL lies in the different ratios of the dust formation radius to the Roche lobe radius, as well as in the varying accretion efficiencies. In cases where this radius ratio was greater than unity, particles did not reach escape velocity, thereby ensuring their transfer to the accretor through $\rm L_{1}$, while for cases with a ratio of less than unity lead to BHL accretion. We note, that despite the differences in the approaches, our results are remarkably similar to theirs.

\citet{2019MNRAS.485.5468I} conducted a study to assess the potential for WDs to reach SNIa conditions via wind accretion, using both the BHL and WRLOF mechanisms. They employed the BHL method to calculate the mass transfer rate through isotropic wind \cite[]{1988A&A...205..155B}, and followed the approach outlined in \citet{2013A&A...552A..26A} to estimate the mass transfer rate via the WRLOF mechanism. Their findings indicated that WRLOF yields substantially higher accretion rates, providing a better chance for WDs to reach SNIa conditions. Our results define the regime in which WRLOF can occur, placing minimal constraints on the conditions under which SNIa may be possible.

We stress that it is unlikely for an actual system to spend its entire mass loss phase in any one type of mass transfer mechanism, as clearly demonstrated in Figure \ref{twoseven}. Moreover, gravity plays a crucial role in shaping the wind, pulling it and preventing it from exhibiting isotropy. Furthermore, in the WRLOF scenario, assuming \textit{all} the wind to be transferred through the $\rm L_1$ point is a significant stretch, as RL lack precise confinement, allowing some of the material to escape. 
Thus, in reality, it is likely that both mechanisms operate to some extent, while the regime limits that we have found indicate the dominant mechanism that is relevant.

Simulations by \cite{2024MNRAS.527.4806V}, where the authors used different separations to study the effects of BHL accretion rate and nova outcomes, resulted in models with smaller separations attaining higher accretion rates  and the WD gaining mass, indicating its potential for growth. However, the accretion mechanism could not provide a sufficient growth rate, and the donor could not supply a sufficient amount of mass to grow the WD to the Chandrasekhar mass, thus these systems are unlikely to be potential SNIa progenitors. This suggests that WRLOF, showing indication of being a more efficient mass transfer mechanism at even smaller separations, should be investigated further.

Since WRLOF gravitationally focuses the wind toward the WD at high mass transfer efficiencies --- $\sim50\%$ of the mass lost by the donor is transferred through the $\rm L_1$ point to the WD \cite[]{2012BaltA..21...88M} --- 
the accretion rate will be higher than that obtained for the BHL mechanism which transfers only a small fraction of the wind to the WD ---  less than 10\% \cite[]{2024MNRAS.527.4806V}
(for a given system and wind rate). This results in more matter being accreted within a given time frame, leading to more frequent eruptions, as higher accretion rates shorten the recurrence time ($ t_{\rm rec}$) \cite[]{1994ApJ...424..319K,1995ApJ...445..789P,2005ApJ...623..398Y,2012BaltA..21...76S,2015MNRAS.446.1924H,2021MNRAS.505.3260H}. A shorter $ t_{\rm rec}$ leads to less mixing (via diffusion and convection) of the accreted matter, resulting in weaker eruptions, as fusion occurs at shallower points on the surface \cite[]{1995ApJ...445..789P,2005ApJ...623..398Y,2022MNRAS.511.5570H,2022MNRAS.515.1404H}. This will eventually lead to the growth of the WD as less matter will be ejected during each eruption \cite[]{1995ApJ...445..789P,2005ApJ...623..398Y,2012BASI...40..419S,2019ApJ...879L...5H}. Moreover, since WRLOF leads to substantially less mass lost to the surroundings, it will have an ample supply of mass (as apposed to the BHL mechanism) to continue providing the WD. Thus, the mass transfer mechanism and accretion rates are closely interconnected, ultimately determining the characteristics of a nova \cite[]{1984ApJ...281..367P,1994ApJ...424..319K,1995ApJ...445..789P,2005ApJ...623..398Y,2012BaltA..21...76S,2018A&A...616L...3B,2018MNRAS.473..747C,2021MNRAS.501..201H,2021MNRAS.505.3260H,2022MNRAS.517.3864S}.

The low mass transfer efficiency of the BHL mechanism fails to grow the WD towards the Chandrasekhar limit, as the WD can only accrete a small portion of the mass within the evolutionary timescale of the donor. A hypothetical solution for substantial WD growth in the BHL scenario would require either a donor with significantly more mass to donate or one that remains on its evolutionary timescale for a longer period. For instance, if we consider a $1.25 M_\odot$ WD accreting from the wind of a an AGB with an initial equal mass at a $10\%$ mass transfer efficiency, only $0.125 M_\odot$ will be transferred. Even if we assume that only half of this mass is ejected in a nova eruption, there is not enough mass to reach the Chandrasekhar mass, and we stress that assuming only half is ejected is highly overestimated based on nova models \cite[]{1995ApJ...445..789P,2005ApJ...623..398Y,2015MNRAS.446.1924H,2016ApJ...819..168H,HILLMAN20201072,2021MNRAS.501..201H,2024MNRAS.527.4806V}. In contrast, in the WRLOF mechanism, direct mass transfer ensures that a substantial fraction (most) of the mass is transferred to the WD, thus, potentially facilitating the growth of the WD and making it a more likely mechanism that can produce SNIa progenitors. This is supported by \cite{2019MNRAS.485.5468I}, where 90\% and 97\% of WDs with masses of 1.25 ${M_\odot}$ and 1.35–1.37 ${M_\odot}$, respectively, in their models reached SNIa. 
However, only self-consistent numerical simulations of such systems with recurrent nova eruptions can determine the fate of such WDs.

\section{Conclusions}\label{Conclusion}
We have explored the entire feasible range of the $ M_{\rm WD}, M_{\rm D}, R_{\rm D}$ parameter space for AGB donors in symbiotic systems and calculated for each parameter combination a separation limit that defines whether the dominant mass transfer mechanism is WRLOF or BHL.
We considered the same mass of AGB across various initial stellar mass models, aiming to encompass the entire possible range of radii associated with a particular AGB mass.
Our results show a correlation between the parameters, in such a way that the limiting separation, $a_{\rm lim}$, increases with the increase of any of the three parameters ($ M_{\rm WD}$, $ M_{\rm D}$ or $ R_{\rm D}$).

We have demonstrated how the dominant mass transfer mechanism in systems with AGB donors that have a binary separation of order the limiting value ($a_{\rm lim}$) could potentially alternate between WRLOF and BHL due to the rapidly changing radius that causes $a_{\rm lim}$ to shift between being smaller or larger than the system's actual separation. Moreover, the mass lost from the system due to substantial wind loss during BHL, less substantial wind loss during WRLOF and even mass ejected due to nova eruptions, all play a role in determining the actual evolving separation, as it affects the total angular momentum of the system. 

It is also important to understand the mass transfer mechanism of the system, as the mass transfer rate, accretion rate, and accretion efficiency play crucial roles in determining the outcome of a nova eruption. A high accretion rate combined with good accretion efficiency favours the growth of the WD, as more matter is accreted, resulting in a shorter recurrence time and less matter being ejected with each eruption. An additional important aspect is that since substantially less mass is lost from the system, the donor might have a sufficient supply of mass to grow the WD to the Chandrasekhar mass, meaning that symbiotic systems in which the donor predominantly transfer mass to the WD via the WRLOF mechanism, might possibly be SNIa progenitors.

In this work, we mapped the parameter regimes of WRLOF and BHL, which is a step towards future understanding of the evolution of binary system  including, the possible evolution of a system from one mass transfer mechanism to another, in particular, but not limited to, the continued study of systems that produce nova eruptions.
 
\section*{Acknowledgments}
We thank an anonymous referee for very helpful comments. We acknowledge support from the Ariel University Research and Development Authority. We acknowledge the Ariel HPC Centre at Ariel University for providing computing resources.

\section*{Data availability}
The data co-related with this work will be shared on reasonable request to the corresponding author.

\appendix

\renewcommand{\thetable}{A1}
\section{Tables }
\begin{table*}
    %\centering
    \hspace*{-1.5cm}
    \begin{tabular}{c c c c c c c }
    %\vspace{0.1cm}
    \hline 
    \multicolumn{1}{c}{}
    \vspace{0.1cm}
      $\rm M_{D}[M_\odot]$ &$\rm R_{D}(\pm0.05)[\mathrm{AU}]$ &$a_{\rm lim}(\pm0.05)[\mathrm{AU}]$ &$a_{\rm lim}(\pm0.05)[\mathrm{AU}]$ &$a_{\rm lim}(\pm0.05)[\mathrm{AU}]$ &$a_{\rm lim}(\pm0.05)[\mathrm{AU}]$ &$a_{\rm lim}(\pm0.05)[\mathrm{AU}]$ \\

       && $\rm M_{WD}=0.6M_\odot$&$\rm M_{WD}=0.75M_\odot$&$\rm M_{WD}=1.0M_\odot$& $\rm M_{WD}=1.25M_\odot$& $\rm M_{WD}=1.4M_\odot$ \\
     \hline
       & 0.5  & 10.3 &10.8    &11.5  &12.2  &12.5 \\
       & 0.8  & 11.2 &11.8    &12.6  &13.3  &13.7\\
       & 1.0  & 11.8 &12.4    &13.2  &14.0  &14.3\\
$0.75\pm0.05$& 1.1  & 12.5 &13.2    &14.1  &14.9  &15.3\\
       & 1.2  & 12.6 &13.3    &14.2  &14.9  &15.4\\
       & 1.3  & 13.0 &13.6    &14.6  &15.4  &15.8\\
       & 1.4  & 13.2 &13.9    &14.9  &15.7  &16.2\\
       \\
     \hline
       & 0.7  & 12.7 &13.4    &14.2  &15.0  &15.4\\
       & 0.9  & 13.7 &14.4    &15.3  &16.1  &16.6\\
       & 1.0  & 14.1 &14.8    &15.8  &16.6  &17.0\\
       & 1.1  & 14.6 &15.3    &16.3  &17.2  &17.6\\
       & 1.2  & 14.7 &15.4    &16.4  &17.3  &17.8\\
$1.0\pm0.05$& 1.3  & 15.0 &15.7    &16.7  &17.6  &18.1\\
       & 1.6  & 15.7 &16.5    &17.6  &18.5  &19.0\\
       & 1.7  & 16.5 &17.3    &18.4  &19.4  &19.9\\
       & 1.8  & 16.7 &17.5    &18.6  &19.6  &20.1\\
       & 1.9  & 16.9 &17.8    &18.9  &19.9  &20.5\\
       & 2.1  & 17.7 &18.6    &19.8  &20.8  &21.4         \\
       \\
      \hline
      %\\
       & 0.8  & 15.0 &15.8    &16.8  &17.6  &18.1 \\
       & 1.0  & 15.6 &16.4    &17.4  &18.3  &18.8\\
       & 1.1  & 16.2 &17.0    &18.1  &19.0  &19.5\\
 $1.25\pm0.05$ & 1.4  & 17.1  &17.9   &19.0  &20.0  &20.5\\
       & 1.5  & 17.3 &18.1    &19.3  &20.2  &20.8\\
       & 1.6  & 17.7 &18.6    &19.8  &20.8  &21.3\\
      & 1.9  & 18.6 &19.4    &20.7  &21.7  &22.3\\
       & 2.2  & 19.6 &20.5    &21.8  &22.9  &23.5\\
       \\
       \hline
     
      %\\
       & 0.8  & 17.1 &17.9    &19.0  &19.9  &20.4 \\
      & 1.0  & 17.7 &18.5    &19.7  &20.6  &21.2\\
       & 1.1  & 18.0 &18.8    &19.9  &20.9  &21.5\\
       & 1.3  & 18.6 &19.5    &20.7  &21.7  &22.3\\
 $1.5\pm0.05$& 1.4  & 18.9 &19.7    &20.9  &22.0  &22.5\\
       & 1.6  & 19.5 &20.4    &21.6  &22.7  &23.3\\
       & 1.7  & 19.8 &20.7    &22.0  &23.1  &23.7\\
       & 1.9  & 20.4 &21.3    &22.6  &23.7  &24.3\\
       & 2.0  & 20.7 &21.7    &23.0  &24.1  &24.8\\
       \\
      \hline
       & 0.8  & 19.0 &19.9    &21.1  &22.1  &22.6 \\
       & 1.0  & 19.5 &20.3    &21.6  &22.6  &23.2\\
       & 1.1  & 19.8 &20.7    &21.9  &23.0  &23.6\\
   $1.75\pm0.05$& 1.3  & 20.5 &21.4    &22.7  &23.8  &24.4\\
       & 1.5  & 20.9 &21.8    &23.1  &24.2  &24.8\\
       & 1.6  & 21.5 &22.4    &23.7  &24.9  &25.5\\
       & 1.8  & 21.9 &22.8    &24.2  &25.4  &26.0\\
     \hline  
    \end{tabular}
    \caption{The limiting separation ($a_{\rm lim}$) for different combinations of WD mass ($ M_{\rm WD}$), AGB mass ($ M_{\rm D}$) and AGB radius ($ R_{\rm D}$).}
    \label{tab:Table new1}
\end{table*}

\begin{table*}
%    \vspace{-3cm}
    %\centering
    \hspace*{-1.5cm}
    \begin{tabular}{c c c c c c c }
    %\vspace{0.1cm}
    \hline 
    \multicolumn{1}{c}{}
    \vspace{0.1cm}
      $\rm M_{D}[M_\odot]$ &$\rm R_{D}(\pm0.05)[\mathrm{AU}]$ &$a_{\rm lim}(\pm0.05)[\mathrm{AU}]$ &$a_{\rm lim}(\pm0.05)[\mathrm{AU}]$ &$a_{\rm lim}(\pm0.05)[\mathrm{AU}]$ &$a_{\rm lim}(\pm0.05)[\mathrm{AU}]$ &$a_{\rm lim}(\pm0.05)[\mathrm{AU}]$ \\
     
       && $\rm M_{WD}=0.6M_\odot$&$\rm M_{WD}=0.75M_\odot$&$\rm M_{WD}=1.0M_\odot$& $\rm M_{WD}=1.25M_\odot$& $\rm M_{WD}=1.4M_\odot$ \\
     \hline
     & 0.7  & 20.4 &21.3    &22.5  &23.6  &24.2 \\
       & 0.9  & 21.1 &22.0    &23.3  &24.4  &25.0\\
       & 1.1  & 21.6 &22.5    &23.8  &24.9  &25.5\\
   $2.0\pm0.05$ & 1.3  & 22.2 &23.2    &24.5  &25.7  &26.3\\
       & 1.4  & 22.7 &23.6    &25.0  &26.2  &26.8\\
       & 1.5  & 23.0 &23.9    &25.3  &26.5  &27.2\\
       & 1.7  & 23.5 &24.5    &26.0  &27.2  &27.8\\
       & 1.9  & 24.0 &25.0    &26.5  &27.7  &28.4\\

       \\
       \hline
        & 0.8  & 22.5 &23.4    &24.7  &25.9  &26.5\\  
        & 0.9  & 22.9 &23.9    &25.2  &26.4  &27.0 \\
        & 1.2  & 23.7 &24.7    &26.1  &27.3  &28.0\\
$2.25\pm0.05$& 1.4  & 24.2 &25.2    &26.6  &27.8  &28.5\\
        & 1.5  & 24.6 &25.6    &27.1  &28.3  &29.0\\
        & 1.6  & 24.9 &25.9    &27.4  &28.6  &29.3\\
        & 1.8  & 25.4 &26.5    &28.0  &29.3  &30.0\\
       
     \\
     \hline
             
        & 0.9  & 24.6 &25.6    &27.0  &28.3  &28.9\\
        & 1.1  & 25.4 &26.4    &27.9  &29.1  &29.8\\
$2.5\pm0.05$ & 1.2  & 25.5 &26.6    &28.0  &29.3  &30.0\\
        & 1.4  & 26.0 &27.0    &28.5  &30.0  &30.5\\
        & 1.6  & 26.7 &27.7    &29.3  &30.6  &31.3\\
        & 1.8  & 27.2 &28.3    &29.9  &31.2  &32.0\\
             
       \\
      \hline
        & 1.0  & 26.6 &27.6    &29.1  &30.4  &31.1\\
$2.75\pm0.05$& 1.2  & 27.3 &28.4    &29.9  &31.3  &32.0\\
        & 1.3  & 27.5 &28.6    &30.2  &31.5  &32.3\\
        & 1.6  & 28.5 &29.6    &31.2  &32.6  &33.3\\
       
       \\
      \hline
            
        & 1.0  & 28.5 &29.6    &31.2  &32.6  &33.3\\
$3.0\pm0.05$& 1.2  & 29.1 &30.2    &31.8  &33.2  &33.9\\
        & 1.4  & 29.6 &30.7    &32.4  &33.8  &34.5\\
        & 1.6  & 30.1 &31.3    &33.0  &34.4  &35.2\\

       \\
      \hline
$3.25\pm0.05$& 1.2  & 30.7 &31.9    &33.5  &35.0  &35.7\\
        & 1.5  & 31.7 &32.9    &34.6  &36.1  &36.9\\
   \\

    \hline   
            
$3.5\pm0.05$& 1.2  & 32.4 &33.6    &35.3  &36.8  &37.6\\
        & 1.4  & 33.0 &34.2    &35.9  &37.4  &38.3\\

    \hline   
    \end{tabular}
    \caption{--- continued.}
    \label{tab:Table new2}
\end{table*}

\section{Parameter Space equation}\label{Apendix}

A detailed derivation of Equation \ref{a_lim} is provided below. 

Rearranging Equation \ref{1} and solving for $R_{\rm RL,lim}$ results in:
\begin{equation} 
    R_{\rm RL,lim} {v_\infty}^2 = GM_D-v_\infty (v_s-v_\infty)R_D+\sqrt{G^2 {M_D}^2-2GM_DR_Dv_\infty (v_s-v_\infty)},
\end{equation}

Reorganizing and integrating Solar units for mass and radii gives:

\begin{equation} 
    \frac{R_{\rm RL,lim}}{R_\odot}  = \frac{G M_\odot M_D}{{v_\infty}^2 M_\odot}\left( 1+ \sqrt{1+\frac{2\frac{R_D}{R_\odot}v_\infty (v_\infty-v_s)}{GM_\odot \frac{M_D}{M_\odot}}}\right)+\frac{v_\infty (v_\infty-v_s)}{{v_\infty}^2}\frac{R_D}{R_\odot}.
\end{equation}

We define a dimensionless constant; 
\begin{equation}
\xi^*=\frac{GM_\odot}{v_\infty^2R_\odot},
\end{equation} 
which, while using CGS units for $G$, $M_\odot$ and $R_\odot$ and adopting $v_\infty\approx20\rm km/s$, yields $\xi^*\approx500$. 

Defining the mass and radii in Solar units allows a more simplified presentation of $R_{\rm RL,lim}$ in the form:

\begin{equation} \label{B3}
    R_{\rm RL,lim}  = \xi^* M_D\left( 1+ \sqrt{1+\frac{2R_D (v_\infty-v_s)}{\xi^*v_\infty{M_D}}}\right)+\frac{ (v_\infty-v_s)}{{v_\infty}}R_D.
\end{equation}

By substituting a dimensionless parameter, $\omega$, for $(v_\infty-v_s)/v_\infty$, which is less than, but of the order of unity, Equation \ref{B3} can be rewritten as:

\begin{equation} \label{B4}
    R_{\rm RL,lim}  = \xi^* M_D\left( 1+ \sqrt{1+\frac{2\omega R_D}{\xi^*{M_D}}}\right)+\omega R_D.
\end{equation}

Reorganizing this gives:

\begin{equation}
 R_{\rm RL,lim}=\sqrt{\xi^* M_{\rm D}}\left(\sqrt{\xi^*M_{\rm D}}+\sqrt{\xi^*M_{\rm D}+2\omega R_{\rm D}}\right)+\omega R_{\rm D}  
\end{equation}

which can be simplified by defining: $\xi=\sqrt{\xi^*M_D}$, to obtain:

\begin{equation} \label{B5}
    R_{\rm RL,lim}  = \xi \left( \xi+ \sqrt{\xi^2+{2\omega R_D}}\right)+\omega R_D,
\end{equation}

This expression for the limiting RL radius may be used in Equation \ref{a_orig} (Eggleton's prescription) to obtain $a_{\rm lim}$:

\begin{equation} \label{B6}
a_{\rm lim} =\left(\xi\left(\xi +\sqrt{\xi^2+2\omega R_D} \right) +\omega R_D\right)  \frac{0.6q^{2/3}+\ln(1+q^{1/3})}{0.49q^{2/3}}
\end{equation}
thus constituting an expression of the four parameter dependency: $a_{\rm lim}(M_{\rm WD}, M_{\rm D}, R_{\rm D})$. 
\begin{figure*}\renewcommand\thefigure{B1}
%\hspace*{-0.6cm}
    \includegraphics[trim={2cm 0cm 0cm 1cm},clip ,width=0.99\columnwidth]{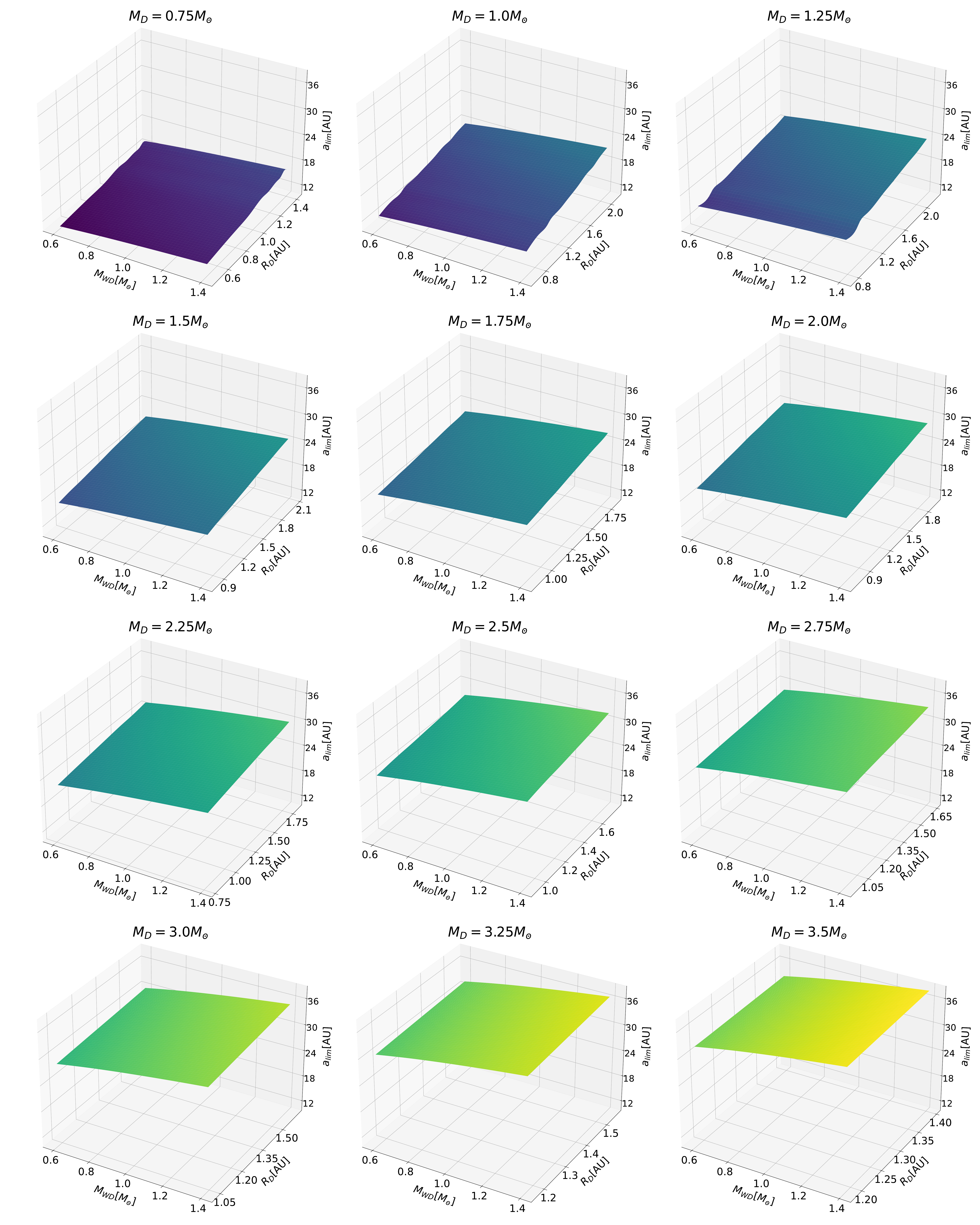}
   \caption{The limiting separation ($a_{\rm lim}$) for each AGB donor mass ($ M_{\rm D}$) model with five WD masses ($ M_{\rm WD}$) in the range $0.6-1.4M_\odot$. Systems that fall below the surface will experience accretion by the WRLOF mechanism, while systems that fall above the surface will result in mass transfer via the BHL mechanism.}
   \label{Separtionsurf}
\end{figure*}

\bibliography{ref}
\bibliographystyle{aasjournal}

\end{document}